\documentclass[a4paper,10pt]{article}
\usepackage{amsmath,amsfonts,amssymb,slashed,cancel,graphicx,epsfig,graphics,hyperref}

\newcommand{\be}{\begin{equation}}
\newcommand{\ee}{\end{equation}}

\title{\textbf{A holographic bound on the scaling contribution to black hole entropy}}
\author{B. Gaasbeek\footnote{bram\_mmm@hotmail.com}}
\date{\textit{Institute for Theoretical Physics, University of Leuven}}
\begin{document}
\maketitle

\abstract{We discuss the existence of scaling solutions for multicenter black hole configurations. One of the central results is the equivalence between the existence of two centered scaling solutions and the holographic entropy bound. This equivalence (and another one) are proved rigorously at the end of the paper, and allow to simplify the process counting of certain (fuzzball-like) contributions to black hole entropy.}

\section{Introduction.}

Multi-centered black hole solutions have been known for a long time \cite{MajPap}. In pure Einstein-Maxwell theory these configurations are just marginally bound (the electrostatic and gravitational forces canceling exactly) so they could be destroyed easily; by throwing in an additional electron for example.
However, the additional scalar fields, electric-magnetic charge pairs and non-linear interactions in supergravity theories allow forming genuine bound states of black holes \cite{DenefFirst}. Indeed, these solutions persist also away from extremality, with some probe examples constructed recently \cite{AnusInc}\cite{BertScoop}.

An interesting role in this story is played by scaling solutions \cite{EntropyFunction}\cite{BenaWarner}. This is the collection of multicenter black hole solutions where the centers can be arranged  closely without actually merging. More precisely, for some solutions the entire constellation is sunk deep into a deep AdS-throat, which make it nearly indistinguishable from a single centered black hole to an outside observer.  
This suggests that at least a part of or the entropy of the corresponding extremal single center black hole must come from these scaling solutions \cite{Dieter}\cite{DGSVY}. So scaling solutions provide an attractive way to obtain (at least a part of) the moduli space of extremal black hole microstates, much along the lines of the fuzzball-proposal \cite{Mathur}.\\
A major obstruction in counting the contribution of these scaling solutions to the black hole index of the total charge (or corrections thereof \cite{Sen}) is their existence and validity. In this paper (based on earlier work \cite{Thesis}) we make some progress in this respect by proving two simple criteria which are equivalent to the full existence of \textit{two}-centered scaling solutions solutions. We also provide some hints for similar conditions for the case of more centers.

\section{Multicenter black holes and scaling solutions}

\subsection{4d multicenter black holes}

We quickly review the basic equations of 4d multicenter black holes in the framework of type II string theory. Each black hole in such a configuration is characterized by its charges under the vector fields in the low energy supergravity theory. One of these vector fields is the graviphoton, the others sit in vector multiplets. We denote these charge vectors by $\Gamma_i$, where $i$ runs from $1$ to $n$, the number of centers, and similarly we denote the locations of the centers by $\vec{x}_i$.

The metric of a stationary (single- or multi-centered) black hole configuration is given by 
\be
ds^2 = -e^{2U}(dt+\omega)^2 + e^{-2U} \,d\vec{x}^2
\label{metric}
\ee
Here $\omega$ is a one-form encoding the intrinsic angular momentum of the solution, coming from the presence of both electric and magnetic charges. 
The warp factor $e^{2U}$ is determined (together with the scalar fields) by the BPS equation 
\be
2 \, e^{-U} \,\textrm{Im}(e^{-i\alpha}\Omega) = -H
\label{BPS}
\ee
where $\Omega$ denotes the (normalised) Calabi-Yau 3-form, and $H$ are harmonic functions
\be
H=\sum_{i=1}^{n} \frac{\Gamma_i}{|\vec{x}-\vec{x}_i |} + h
\label{harmonics}
\ee
The constant term in the harmonic functions is determined by the total charge and the moduli at infinity. For the asymptotically flat case
\be
h = - 2 \,\textrm{Im}(\bar{Z}(\Gamma)\Omega)|_{r=\infty}
\ee
with $Z(\Gamma)$ the central charge of the total charge $\Gamma$ ($= \sum_i \Gamma_i$) both evaluated at spatial infinity ($r=\infty$) and thus dependent at the moduli there. The actual positions of the centers are constrained by the \textit{integrability conditions}, 
\be
\sum_{j\neq i}\frac{\langle\Gamma_i, \Gamma_j\rangle}{r_{ij}} = 2 \,\textrm{Im} (\bar{Z}(\Gamma)Z(\Gamma_i))|_{r=\infty} \,\,\,\,\forall i
\label{IC}
\ee
where $r=\infty$ denotes spatial infinity again, and $\langle\,,\rangle$ is the antisymmetric intersection product.

\subsection{Entropy function}
It is clear that BPS equation (\ref{BPS}) is a rather implicit expression for $U$. One way to move forward involves introducing an additional function $S(\Gamma)$, the \textit{entropy function}: 
\be
S(\Gamma) = |Z(\Gamma,t^*(\Gamma)|^2
\ee
where $t^*(\Gamma)$ are the attractor-values \cite{Attractor} 
of the moduli at the horizon of an extremal black hole with charge $\Gamma$.  (Note that $S$ is a quadratic function of the charges, since the central charge $Z$ is linear in terms of $\Gamma$. This property will turn out to be important in what follows.) It can be shown (see f.e. \cite{DenefBatesFlow}) that taking 
\be
e^{-2U} = S(H)
\label{entropyfunction}
\ee
solves the BPS equations. Also, by looking at the near-horizon behavior of the metric (and using linearity of the central charge in terms of $\Gamma$)
one sees that the Bekenstein-Hawking entropy $S^{BH}$ of a center with charge $\Gamma_i$ equals
\be
S^{BH}_i = \pi S(\Gamma_i)
\ee
which justifies the naming of the function $S$. In what follows, we will simply refer to $S$ as \textit{the} black hole entropy - with the proportionality factor $\pi$ understood but neglected.

\subsection{Scaling solutions}

Now we ask how we can arrange these centers so that they are encompassed by an AdS-throat which looks like the near-horizon region of a single black hole. By inspecting (\ref{harmonics}), it is clear that placing the centers very close to each other in coordinate space (so that the $\vec{x}_i$ are nearly identical) will produce a harmonic function which looks like $H\approx \Gamma/r +h$ troughout most of the space. The corresponding geometry will thus be very similar to a single black hole carrying the total charge $\Gamma$ for an outside observer. Only very close to the centers $\vec{x}_i$ will the harmonic function $H$ (and thus the geometry) be distinguishable from that of a single black hole.

\begin{figure}
\begin{center}
\includegraphics[width=10cm,angle=0]{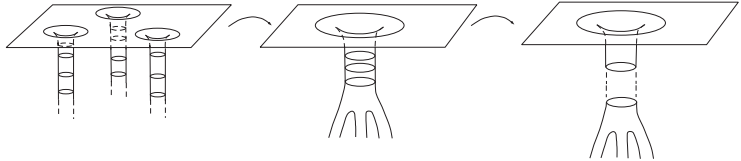}
\caption{\label{fig:down_the_throat} A scaling solution: for some charges, it is possible to move the centers of a multicenter black hole together, up to the point where their near-horizon geometries disappear behind an encompassing AdS$_2$ $\times$ S$^2$ region. This geometry looks like a single extremal black hole for an outside observer (`exterior geometry') but inside sits a multicenter black hole solution living in an asymptotically AdS$_2$ $\times$ S$^2$ space (`interior geometry').}
\end{center}
\end{figure}

By taking the limit of `closely arranged centers' more carefully one can check that the physical distance between the centers doesn't decrease to zero as one might expect naively. An observer who stays near the centers (while they are moved closer together in coordinate space) will always see them as separate black holes, while the asymptotic geometry starts looking like the near-horizon region of a single black hole - an increasingly deep encompassing AdS$_2$ $\times$ S$^2$  throat is forming. This process is shown in figure \ref{fig:down_the_throat}.

Let us consider again the integrability conditions, which basically dictate the distances between the different centers, given their charges and the asymptotic moduli. Looking for solutions where the centers are close to one another in coordinate space means taking
\be
r_{ij} = \lambda \tilde{r}_{ij}, \,\,\lambda\rightarrow 0
\ee
where $r_{ij} = |\vec{x}_{i} - \vec{x}_j|$ is the coordinate distance between two centers.

Plugging the above in the integrability equations (\ref{IC}), multiplying both sides by $\lambda$ and taking the limit, we get
\be
\sum_{j\neq i}\frac{\langle\Gamma_i, \Gamma_j\rangle}{r_{ij}} = 0 \,\,\,\,\,(\forall i)
\label{IC2}
\ee
We can also get to this equation via a different road. We mentioned above that moving black hole centers together creates a common AdS throat around the whole. The means the result can -in the limit of an infinitely long throat- be approached by one consisting of two patches: an asymptotically flat region, with a single black hole throat (with AdS$_2$ $\times$ S$^2$ geometry) and glued inside this, the geometry of a multicenter black hole living in a universe of which the asymptotic region is AdS$_2$ $\times$ S$^2$.\footnote{The inner geometry thus interpolates between one asymptotic AdS region and several AdS near-horizon regions around the centers, a geometry related to the notion of AdS fragmentation and Brill instantons \cite{Brill}\cite{Fragmentation}}
For a multicenter black hole solution with such asymptotics the left hand side of (\ref{BPS}) is zero at spatial infinity, so the harmonic functions have no constant term $h$ and the right hand side of the integrability conditions becomes zero:
\be
H=\sum_{i=1}^{n} \frac{\Gamma_i}{|\vec{x}-\vec{x}_i |}
\ee
and the integrability conditions indeed reduce to \ref{IC2}. A first observation is that equation (\ref{IC2}) allows linear rescaling of the distances $r_{ij}$. So once the centers are close enough, further downscaling proceeds linearly. Doing so just stretches the common throat further. Or from the inner geometry point of view: multicenter black holes in an asymptotically AdS$_2$ $\times$ S$^2$ universe always allow linear scaling up of the positional configuration. This symmetry is referred to as the scaling symmetry. 

For two center scaling solutions, the only requirement is $\langle \Gamma_1,\Gamma_2\rangle=0$, leaving all $r_{12}$ allowed. For three centers the solutions are given by $r_{ij}=|\langle \Gamma_i,\Gamma_j\rangle|$ and all linear rescalings thereof. (Provided that $\langle\Gamma_1,\Gamma_2\rangle$ and its two cyclic permutations are all three either positive or all negative.)

For more than three centers, more positional freedom arises. Indeed, of the $n(n-1)/2$ different distances, $3n-6$ are independent (since the first 3 centers have 3 independent distances amongst them, and one only needs to specify the distances of a center with respect to the first three centers in order to locate it) so the above equations (of which $n-1$ are independent) allows a $2n-5$ dimensional solution space. This space will be cut down by the discrete requirement that the triangle inequalities be satisfied for each subset of 3 centers, but this leaves the dimensionality of the solution space unaffected. 

This suggests the configuration space of scaling solutions can actually be quite large and may very well have nontrivial contributions to the entropy of a single centered black hole carrying the total charge.

\subsection{Existence criteria}

An important caveat is that the entropy function defined higher above need not be defined for every charge. That is: not every charge vector has a supersymmetric solution, hence not all charges \textit{have} attractor moduli. In some specific examples \cite{DenefBatesFlow} 
$S$, is the square root of a discriminant function which turns negative in some regions, which thus lie outside the domain of $S$. Also we expect in several other cases that the entropy function $S$ vanishes on the boundary of its domain, so zero entropy solutions separate valid from singular solutions. This domain may still form a rather complicated region in charge space, even in simple examples.

The relevance of the domain of the entropy function can be seen by writing the metric more explicitly by combining (\ref{metric}) and (\ref{entropyfunction})
\be
ds^2 = - \frac{(dt+\omega)^2}{S(\vec{x})} + S(\vec{x}) \, d\vec{x}^2
\ee
with $S$ depending on $\vec{x}$ via the harmonic functions:  $S(\vec{x})=S(H(\vec{x}))$. This implies that the harmonic function $H$ has to stay inside the domain of $S$ throughout the entire space in order to have a valid geometry. 

Verifying this is a nontrivial task, but crucial in order to be able to determine whether a scaling solution exists, and hence check in general to what extent scaling solutions make up the entropy of extremal black holes.

In this work we make a first but important step in answering that question: in the next sections we will propose and prove two easy criteria which are equivalent to this hard question of existence, for the case of two centered scaling solutions.

\section{Existence of two center scaling solutions}

We now restrict our attention to two-center scaling solutions. These correspond to two charges $\Gamma_1$, $\Gamma_2$ satisfying the integrability condition (\ref{IC2}): 
\be
\langle \Gamma_1,\Gamma_2\rangle=0
\label{ICSS}
\ee
So two charges allow a scaling solution if they are mutually local, i.e. if they have a vanishing intersection product. The first reason for this restriction to two centers is simplicity, even though we will see that even for these simple cases, a lot of interesting things can be said already.

A second reason is the possibility is the fact that two-center scaling solutions and recursive combinations thereof (a throat splitting further in two throats etc) might in some cases constitute a sufficiently rich configuration space to obtain relevant contributions to the partition function \cite{Miranda}, a point we will come back to in the conclusion.

\subsection{Holographic principle}

The most straightforward criterion for the existence of scaling solutions is the holographic principle \cite{Susskind}. Since each scaling solution is enclosed by a surface of area $A = 4 S^{BH}(\Gamma)$ (the area of the sphere of the encompassing throat) the holographic principle dictates that sum of the individual entropies is bounded by this area. In terms of the entropy function $S$ (which differs just by a factor of $\pi$ from the actual black hole entropy)
\be
 \sum_{i=1}^n S(\Gamma_i) \leq S(\Gamma)
\ee
In the specific case of two centers,
\be
S(\Gamma_1) +S(\Gamma_2)\leq S(\Gamma_1 + \Gamma_2)
\ee
This provides us at least with a necessary condition on the existence of the scaling solution.

\subsection{Scaling solution (non)flowtree}
Other quantities of interest are the central charges of the two centers. For regular multi-center black holes, the split attractor flow conjecture \cite{SAFC} relates the existence of a solution to the possibility of dynamically creating it from its basic constituents. This  is done by changing the asymptotic moduli across a wall of marginal stability in moduli space, which separates/joins two clusters. For two-center solutions, this can be done at asymptotic moduli which align the central charges of the two centers:\footnote{A short derivation of the locus of the  moduli at which such decay can occur goes as follows. Say a multi-center configuration ceases to exist when tuning the asymptotic moduli across some critical value of the moduli. The initial mass is given by $M_{i}=|Z(\Gamma_1+\Gamma_2)|$ with $\Gamma_{1,2}$ the charges of the two clusters. On the other hand the final mass is $M_f= |Z(\Gamma_1)| + |Z(\Gamma_2)|$. The triangle inequality implies $M_i \leq M_f$ while on physical grounds we need $M_i\geq M_f$. Hence we need the equality $|Z(\Gamma_1+ \Gamma_2)| = |Z(\Gamma_1)|+|Z(\Gamma_2)|$ which implies the central charges need to be aligned on the complex plane.}
\be
\textrm{arg} \, Z(\Gamma_1,t_{r=\infty}) = \textrm{arg} \, Z(\Gamma_2,t_{r=\infty}), \,\, \textrm{i.e.}
\ee
\be
Z(\Gamma_1)\bar{Z}(\Gamma_2) |_{t_{r=\infty}} \,\,\,\,\, \textrm{real and positive}
\ee
Alas, this idea can not be repeated for scaling solutions since their scaling symmetry renders any decay process impossible. 
However, it seems plausible that a special role is still played by this quantity, which is also related to the binding energy between the two centers. (The action of a supersymmetric test particle with charge $\Gamma_t$ in an electromagnetic background $\cal{A}$ is given by $S=-\int |Z(\Gamma_t)| ds + c_0 \int \langle\Gamma_t, \begin{cal}A\end{cal}\rangle$, with $c_0$ a numerical constant. For a supersymmetric background, the potential term stemming from this action is $c_0\langle \Gamma_t,\begin{cal}A\end{cal}\rangle  = e^{U} \textrm{Re} (\bar{Z} Z(\Gamma_t))$ with $Z$ the central charge of the background solution. Hence aligned central charges are related to an attractive potential between charges, which is clearly necessary in order to have a bound state.)

\subsection{Proposal}
\label{proposal}
In the two previous sections, we suggested both a requirement (holographic principle) and a consequence (relative orientation of central charges) of the existence of a two center scaling solution. Extensive numerical investigation \cite{Thesis} suggests that a stronger statement is actually true: both criteria are fully equivalent with the existence of a two center scaling solution - i.e. positivity of the entropy function everywhere. This preliminary investigation leads to hypothesize a full equivalence between the following three statements:\footnote{We will use $S$, $S_1$ and $S_2$ as shorthand for  $S(\Gamma)$, $S(\Gamma_1)$ and $S(\Gamma_2)$. Similarly $Z$, $Z_1$, $Z_2$ denote the central charge of $\Gamma$, $\Gamma_1$, $\Gamma_2$; the moduli at which these central charges should be evaluated will be indicated with subscripts.}\footnote{Since the naming of $\Gamma_1$ and $\Gamma_2$ is arbitrary, logical consistency of course requires that (I) is also equivalent to $Z_1\bar{Z}_2|_{t^*(\Gamma_2)}>0$}
\begin{eqnarray}
&Z_1\bar{Z}_2|_{t^*(\Gamma_1)}>0\nonumber&\,\,\,\,\,\, \textrm{(I)}\\
&\Updownarrow\nonumber&\\
& S(\vec{x})\,\, \textrm{positive everywhere}\nonumber&\,\,\,\,\,\, \textrm{(II)}\\
&\Updownarrow\nonumber&\\
&S>S_1+S_2&\,\,\,\,\,\, \textrm{(III)}\nonumber
\end{eqnarray}
under the working assumptions that:
\begin{itemize} 
\item the single-center black hole solutions of the individual centers and total charge exist,  so $S_1>0$, $S_2>0$ and $S>0$. 
\item the entropy function is `typical': it goes to zero on the boundary of its domain.
\end{itemize}
The above equivalences gives us two easy-to-check criteria for the full existence of two-center scaling solutions. We will rigorously prove these equivalences in the next section.

Before doing so, we digress a bit on the last working assumption. We do not have a simple argument that this should always be the case, but it seems to be true in quite some situations. For a cubic prepotential in type IIA for instance, one can argue that this is always the case. The attractor point of a charge $\Gamma$ is unique in these theories, so in order to vary $\Gamma$ out of Dom S, one should make this critical point $t^*$ of $Z(\Gamma,t)$ vanish. The only way to achieve this is to move $t^*$ to the edge moduli space. In appendix A of \cite{Greene} it is shown that $|Z(\Gamma,t)|$ is bounded from below by $|Z^*(\Gamma)|$ times the length of the attractor flow connecting $t$ and $t^*(\Gamma)$. If $t^*$ is pushed off to the edge of moduli space (which in these theories lies at infinite distance) then $|Z(\Gamma,t)|$ would diverge (which is physically not allowed), unless $|Z(\Gamma,t^*)|$ is zero. This suggests that in these theories, the entropy function $S=|Z^*|^2$ has to go to zero on the boundary of its domain. 
We do not have a more general argument for this property to be true, but -as said - it seems to be the case in quite some situations, and we will need it in order to be able to prove the above proposal.

\section{Proof of proposed existence criteria}
In the following section, we prove the equivalences proposed above in a somewhat technical fashion. Readers who are more interested in the physical interpretation, are advised to jump to the conclusion in the next section, where the result and assumptions are again summarized, and where a more physical interpretation is given.
\subsection{ (I) $\Leftrightarrow$ (II)}
First recall \cite{DenefBatesFlow} 
that for every charge $\Gamma\,\in \,\textrm{Dom}\,S$ 
\be
2 \, \textrm{Im} (\bar{Z}^*(\Gamma)\, \Omega^*(\Gamma)) = -\Gamma
\ee
where we use an asterisk to denote that a quantity is evaluated at the attractor moduli of the charge, so $Z^*(\Gamma)=Z(\Gamma,t^*(\Gamma))$ and $\Omega^* = \Omega(t^*(\Gamma))$. If we now apply the above identity to a charge $a\Gamma_1+b\Gamma_2$ (with $a$ and $b$ real and positive) 
and take the intersection product of the equation with $\Gamma_1$, we get
\be
2\, \textrm{Im} \left[\bar{Z}^* (a\Gamma_1+b\Gamma_2)\, Z(\Gamma_1,t^*(a \Gamma_1 + b\Gamma_2))\right] =0
\ee
The right hand side is zero by antisymmetry of the intersection product $\langle\,,\rangle$ and because of the integrability condition (\ref{ICSS}).
Using linearity of the central charge, the previous equation implies 
\be
\textrm{Im} (\bar{Z}_1 Z_2)|_{t^*(a \Gamma_1+b\Gamma_2)} =0
\ee
Hence $Z_1^* \bar{Z}_2|_{t^*(a \Gamma_1+b\Gamma_2)}$ is always real - provided $a \Gamma_1+b\Gamma_2 \in \textrm{Dom}\,S$. Also, because the central charge reaches a minimum at the attractor moduli \cite{Attractor}, we have
\be
|Z_1|^2_{t^*(a \Gamma_1+b\Gamma_2)}\geq |Z_1|^2_{t^*(\Gamma_1)} = S_1 \geq 0
\ee
The last step is true by virtue of the first working assumption (section \ref{proposal}). A similar expression holds for $|Z_2|^2$, so the norm of the (real) quantity $\bar{Z}^* (\Gamma_1) Z(\Gamma_2)|_{t^*(a \Gamma_1+b\Gamma_2)}$ is bounded from below by $\sqrt{S_1 S_2}$, and we arrive at the following property:
\begin{quote}\textbf{Property:} For mutually local charges $\Gamma_1$ and $\Gamma_2$, the quantity $\bar{Z} (\Gamma_1) Z(\Gamma_2)|_{t^*(a \Gamma_1+b\Gamma_2)}$ is real and has the same sign throughout each connected patch of Dom S (as one varies $a$ and $b$) 
\end{quote}
This property will soon prove to be useful. On each point of space, the harmonic function of a two center scaling solution precisely has the form $H=a\, \Gamma_1 + b \, \Gamma_2$ with $a$ and $b$ real and positive. Hence the entropy function is given by
\begin{eqnarray}
S(H) &=& |Z(H,t^*(H))|^2\\
&=& a^2 |Z(\Gamma_1,t^*(H))|^2 +  b^2 |Z(\Gamma_2,t^*(H))|^2 + 2ab\,\textrm{Re} (Z_1 \bar{Z}_2)|_{t^*(H)}
\label{ExpandedEntropy}
\end{eqnarray}
We can now argue that if $Z_1 \bar{Z}_2|_{t^*(\Gamma_1)}>0$, the entire line segment $[\Gamma_1,\Gamma_2]$ in charge space lies in Dom S. By the working assumption $S_1>0$, so the above expression is strictly positive for $a=1$, $b=0$. Now start varying $a$ and $b$ so that $a\Gamma_1+b\Gamma_2$ runs over the line segment $[\Gamma_1,\Gamma_2]$ in charge space. By the earlier finding, $Z_1 \bar{Z}_2|_{t^*(a\Gamma_1+b\Gamma_2)}$ will stay positive, (given that is was so initially) as long as we stay inside Dom S. Should we reach the boundary of that domain, $S(a\Gamma_1+b\Gamma_2)$ should go to zero, by the second working assumption (section \ref{proposal}). But the right hand side of the above equation is strictly positive, even should $Z_1\bar{Z}_2$ decrease to zero, as the first two terms are bound from below by $a^2 S_1 + b^2 S_2$. Hence $S$ can impossibly become zero on the line segment $[\Gamma_1,\Gamma_2]$ in charge space.
Because of quadratic dependence of $S$ on the charges, this implies the entire image of $H$ (which comprises charges of the form $a \Gamma_1 + b \Gamma_2$ with $a$ and $b$ positive) lies inside Dom S.
\textbf{This proves (I) $\Rightarrow$ (II).}

We now show (II) $\Rightarrow$ (I). So suppose (II) is true and (I) is not, we'll arrive at a contradiction. 
Take a charge $\Gamma_a$ running along the interval $[\Gamma_1,\Gamma_2]$, for example $\Gamma_a=a\Gamma_1+(1-a) \Gamma_2$, $a$ ranging from 0 to 1. 
Consider the function 
\be
f: a \rightarrow Z_a \bar{Z}_2|_{t^*(\Gamma_a)}
\ee
with $a$ ranging from 0 to 1. The initial value $f(0)$ is strictly positive: $f(0)=|Z_2^*|^2 = S_2 >0$. On the other hand $f(1)$ is negative: $f(1)=Z_1 \bar{Z}_2|_{t^*(\Gamma_1)} \leq 0$ because we supposed (I) to be false. Still holding on to (II), $f$ is defined along the entire interval, so there must be an $\tilde{a}\in[0,1]$ such that $f(\tilde{a})=Z_{\tilde{a}}\bar{Z}_2|_{t^*(\Gamma_{\tilde{a}})}=0$. Because of the attractor mechanism $|Z_2|_{t^*(\Gamma_{\tilde{a}})}>|Z_2|_{t^*(\Gamma_2)}=\sqrt{S_2}\neq0$ whence $Z_{\tilde{a}}\bar{Z}_2|_{t^*(\Gamma_{\tilde{a}})}$ can only be zero if $\bar{Z}_{\tilde{a}}|_{t^*(\Gamma_{\tilde{a}})}$ and $S(\Gamma_{\tilde{a}})=0$. This contradicts the initial assumption (II). \textbf{So we have proven (II) $\Rightarrow$ (I).}

\subsection{(II) $\Leftrightarrow$ (III)}

We will start off by showing (II) $\Rightarrow$ (III). This is not too hard actually. Re-using (\ref{ExpandedEntropy}) with $a=b=1$, we get\footnote{The real part is automatic due to the property shown above.}
\be
S= |Z_1|^2_{t^*(\Gamma)} + |Z_2|^2_{t^*(\Gamma)} + 2 Z_1 \bar{Z}_2 |_{t^*(\Gamma)}\geq S_1 + S_2 + 2 \,Z_1 \bar{Z}_2 |_{t^*(\Gamma)}
\ee
In order to arrive at (III), we would like to argue that the last term is positive. This is indeed the case. From the equivalence shown above, we know that (II) implies $ Z_1 \bar{Z}_2 |_{t^*(\Gamma_1)}>0$ and because the entire interval $[\Gamma_1,\Gamma_2]$ is in dom S, we can conclude $Z_1 \bar{Z}_2 |_{t^*(\Gamma/2)} = Z_1 \bar{Z}_2 |_{t^*(\Gamma)} > 0$. 
\textbf{We have thus shown (II) $\Rightarrow$ (III)}.

\begin{figure}
\begin{center}
\includegraphics[width=8cm,angle=0]{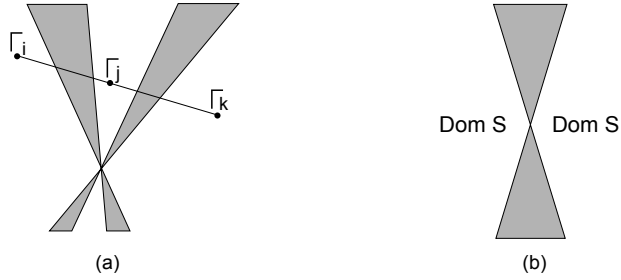}
\caption{\label{fig:domains} Regions not in Dom S are shown in grey. (a) If we suppose there are several sectors lying outside Dom S, we can pick charges
$\Gamma_i$, $\Gamma_j$ en $\Gamma_k$ as on the image. This necessarily raises contradictions. (b) This means there is at most one sector (on both sides of course) which doesn't belong to Dom S.}
\end{center}
\end{figure}

Before proceeding to prove the converse, we first obtain another useful property. We want to investigate the possible shape of Dom S on a plane in charge space spanned by two mutually local charges. First of all, the quadratic dependence of S on charges implies the domain consists of one or more `slices of pie' around the origin, as shown on figure \ref{fig:domains} (a). In what follows we will show it necessarily has a simple shape; as shown on figure \ref{fig:domains} (b). To show this, suppose Dom S (restricted to one particular plane in charge space) consists of more than two pieces. Choose charges $\Gamma_i$, $\Gamma_j$ and $\Gamma_k$ as on \ref{fig:domains} (a). None of the indicated intervals stays within Dom S entirely, while the endpoints do. The equivalence (I) $\Leftrightarrow$ (II) shown above then implies \footnote{Strictness of the inequalities below follows from the fact that $\Gamma_{i,j,k}$ all have positive entropy and that the norm of the central charge is minimal on the attractor values. If $Z_i|_{t^*(\Gamma_j)}$ would be zero for instance, this necessarily means $|Z^*(\Gamma_i)|\leq0$ and thus $S_i=0$, which cannot be the case under the current working assumptions.} \footnote{Since the three charges lie inside Dom S they all have positive entropy. The total charge of any pair does not necessarily have positive entropy however. Even though this was a working assumption, one can verify that $S>0$ was not needed to prove (I)$\Leftrightarrow$(II). So we are allowed to apply that property to each pair of the charges $\Gamma_i$, $\Gamma_j$ and $\Gamma_k$.}
\be
Z_i \bar{Z}_j|_{t^*(\Gamma_j)}<0 \,\,\,\,\textrm{and}\,\,\,\,\, Z_j \bar{Z}_k|_{t^*(\Gamma_j)}<0 \,\,\,\,
\ee
Multiplying these inequalities, and using $Z_j \bar{Z}_j|_{t^*(\Gamma_j)}=S_j>0$, we get
\be
Z_i \bar{Z}_k|_{t^*(\Gamma_j)}>0
\ee
Using $\Gamma_j=a\Gamma_i+b\Gamma_k$ with $a$ and $b$ positive, the last inequality implies
\be
Z_i \bar{Z}_j|_{t^*(\Gamma_j)} = a Z_i \bar{Z}_i|_{t^*(\Gamma_j)}  + b Z_i \bar{Z}_k|_{t^*(\Gamma_j)} >0
\ee
This contradicts the inequality $Z_i \bar{Z}_j|_{t^*(\Gamma_j)}<0$ above, so we conclude: 
\begin{quote}On a plane spanned by mutually local charges, at most one slice can be cut out of Dom S (on both sides, of course) like on figure \ref{fig:domains} (b).\end{quote}
Actually, noting that $Z(\Gamma_1)\bar{Z(\Gamma_2)} = - Z(- \Gamma_1)\bar{Z(\Gamma_2)}$ we see that line segments $[\Gamma_1,\Gamma_2]$  and  $[-\Gamma_1,\Gamma_2]$ can not both be entirely in dom S. So we can refine the conclusion above: there is always \textit{precisely} one slice cut out of Dom S, like on \ref{fig:domains} (b).
For what follows, we'll also need a lower bound on the entropy function, on its domain. First we note that for every $0<a<1$
\begin{eqnarray}
\left\{[\Gamma_1,\Gamma_2] \subset \textrm{Dom(S)}\right\} &\Rightarrow& \left\{[a\Gamma_1,(1-a)\Gamma_2] \subset \textrm{Dom(S)}\right\} \\
&\Rightarrow & \left\{S(a \Gamma_1 + (1-a)\Gamma_2)> a^2 S_1 + (1-a)^2 S_2\right\}
\end{eqnarray}
The first step uses homogeneity of $S$, the second applies the -just proven- relation (II) $\Rightarrow$ (III) to charges $a\Gamma_1$ and $(1-a)\Gamma_2$.

\begin{figure}
\begin{center}
\includegraphics[width=5cm,angle=0]{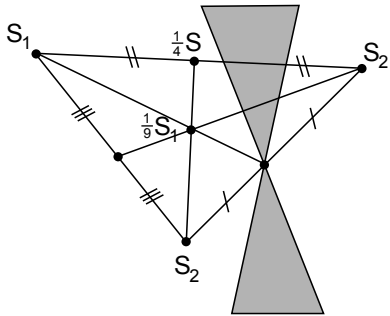}
\caption{
\label{fig:triangle} If the interval $[\Gamma_1,\Gamma_2]$ is not fully in Dom S, we necessarily are in a situation as pictured. The value of the entropy function is shown here, for charges $\Gamma_1$, $\Gamma_2$, $\frac{\Gamma_1+\Gamma_2}{2}$, $-\Gamma_2$ and $\Gamma_1/3$ (barycenter of the triangle).}
\end{center}
\end{figure}

We now have all ingredients to show the remaining implication, (III) $\Rightarrow$ (II). We will do this by arguing $\left\{\textrm{$S(H)$ not positive everywhere}\right\}\Rightarrow\left\{S_1 + S_2 > S\right\}$. The fact that $H$ is not fully contained in Dom S implies we are in a situation like on figure \ref{fig:triangle}. The point corresponding to $-\Gamma_2$ has also been drawn on that figure, and the values of the entropy function for each of these points. From the image, it follows that the entropy $S_b$ at the barycenter of the triangle is $\frac{1}{9} S_1$. On the other hand, applying the lower bound on the entropy function (shown just above) to the interval $[\Gamma/2,-\Gamma_2]$ gives
\be
S_b \geq \frac{4}{9}\left(\frac{1}{4} S\right) + \frac{1}{9} S_2 = \frac{1}{9} (S+S_2)
\ee
Hence 
\be
S_1 \geq 9 \frac{1}{9} (S+S_2) 
\ee
and because $S_2>0$
\be
S_1 + S_2 > S
\ee
\textbf{This shows (III) $\Rightarrow$ (II).}

\newpage
\section{Conclusion, outlook}

We've proven that the existence of a two-center scaling solution is equivalent to
satisfying the holographic principle and also to having the attractor moduli of the total charge positioned on a wall of marginal stability of the two charges:
\be
\left\{Z_1\bar{Z}_2|_{t^*(\Gamma)}>0\right\} \Leftrightarrow \left\{S(H)\,\,\, \textrm{positive everywhere}\right\} \Leftrightarrow \left\{S>S_1+S_2\right\}
\ee
Asumptions to arrive at these conclusions were:
\begin{itemize}
\item Each of the two charges allows a true black hole solution: $S_1>0$ and $S_2>0$.
\item The exterior geometry exists ($S>0$) and this single-black hole solution is compatible with the moduli at infinity. (Single flow between $t(r=\infty)$ and $t^*(\Gamma)$ exists.)
\item The entropy function is continuous on its domain and vanishes on the boundary of its domain. (The validity of this working assumption in quite some situations was argued at the end of section \ref{proposal}.)
\end{itemize}
Especially the second equivalence has a nice physical interpretation: every two-centered scaling solution which (holographically) `fits' into the near-horizon geometry of a single black hole carrying the total charge will actually exist.

It would be very interesting to extend these results to more general scaling solutions. Regarding the holographic principle, it is instructive to consider for example the three charges $\Gamma_0$, $\Gamma_0$, $-\Gamma_0/3$. They do obey the holographic principle (and the scaling solution integrability conditions) but there will clearly be a point where $H=0$ so the solution is ill. This implies straightforward extrapolation of our results to more centers will not work. One may consider imposing the holographic principle on each partition of the charges, which would at least rule out examples like this. Whether such a criterion is sufficient (or necessary) isn't immediately clear though.

Also, it may be fruitful to explore the extent to which the other criterion could be generalized. A hands-on approach is to investigate numerically the locus of the flow tree (here understood to mean the image of the moduli when seen as functions of the coordinates) for larger scaling solutions, and its relation to the relevant walls of marginal stability. Also, the relative orientation of central charges might be relevant. Though they are not necessarily parallel anymore for more centers (as the integrability conditions no longer require the charges to be mutually local) they could still be demanded to point in the same direction for example, such as (and) asking $\textrm{Re}\, Z_i\bar{Z}_j>0\,\, \forall\, i,j$ for instance. 

Lastly, it would be very interesting to count for some concrete example the number of binary splittings $\Gamma \rightarrow \Gamma_1 + \Gamma_2$ one can make. (Using one of the two existence criteria proved above to test validity of the split solution, without actually having to construct it.) This in principle very simple calculation could give an idea of contribution the entropy of the single center black hole receives from scaling solutions for some concrete example. 
The physical idea of course is that one could iterate this process (applying it to $\Gamma_1$ and $\Gamma_2$ etcetera) to create a binary splitting, all the way up to elementary particles. The complete tree-like geometry one would end up with would then be free of horizons, and would (for the cases where the number of binary splittings matches the entropy of the total charge) give an attractive interpretation of the internal degrees of freedom of the black hole visible to an external observer, much along the lines of the fuzzball proposal.

\section{Acknowledgements}
The author wishes to thank D. Anninos and B. Vercnocke for many useful comments on the draft. The author is also greatly indebted to F. Denef, not only for proposing the problem but also for many discussions and providing several crucial points of the proof.

\end{document}